\documentclass[a4paper,11pt]{article}
\usepackage{amsfonts}

\newlength{\dinwidth}
\newlength{\dinmargin}
\def\nn{\nonumber}
\def\non{\nonumber\\}
\def\be{\begin{equation}}
\def\ee{\end{equation}}
\def\ben{\begin{displaymath}}
\def\een{\end{displaymath}}
\def\ba{\begin{eqnarray}}
\def\ea{\end{eqnarray}}

\def\3{\ss}

\def\C{\Gamma}

\def\d{\delta}
\def\e{\varepsilon}

\def\l{\lambda}

\def\p{\pi}

\def\r{\rho}
\def\s{\sigma}
\def\t{\tau}
\def\th{\theta}

\def\w{\omega}
\def\W{\Omega}

\def\C{\mathbb{C}}
\def\Z{\mathbb{Z}}

\def\tpi{\textstyle{\frac1{2\pi i}}}
\def\half{\textstyle{\frac12}}

\makeatletter
\@addtoreset{equation}{section}
\makeatother

\begin{document}
\renewcommand{\thefootnote}{\fnsymbol{footnote}}
\begin{flushright}
DESY 95-202\\
hep-th/9511087\\
November 1995\\
\vspace*{0.7cm}
\end{flushright}
\begin{center}
{\huge On the quantization of isomonodromic deformations on}\smallskip\\
{\huge the torus}\bigskip\\
{\large D.A. Korotkin\footnote{On leave of absence from Steklov
Mathematical Institute, Fontanka, 27, St.Petersburg 191011 Russia} and
J.A.H. Samtleben\bigskip\medskip\\ { II. Institut f\"ur Theoretische
Physik, Universit\"at Hamburg,}\\ { Luruper Chaussee 149, 22761
Hamburg, Germany}}\\ \small E-mail: korotkin@x4u2.desy.de,
jahsamt@x4u2.desy.de
\end{center}
\renewcommand{\thefootnote}{\arabic{footnote}}
\setcounter{footnote}{0}
\vspace*{0.7cm}
\hrule
\begin{abstract}
The quantization of isomonodromic deformation of a meromorphic
connection on the torus is shown to lead directly to the
Knizhnik-Zamolodchikov-Bernard equations in the same way as the
problem on the sphere leads to the system of Knizhnik-Zamolodchi\-kov
equations. The Poisson bracket required for a Hamiltonian formulation
of isomonodromic deformations is naturally induced by the Poisson
structure of Chern-Simons theory in a holomorphic gauge fixing. This
turns out to be the origin of the appearance of twisted quantities on
the torus.\medskip

\end{abstract}
\hrule
\vspace*{1cm}

\section{Introduction}
The fundamental role of the Knizhnik-Zamolodchikov (KZ) equations is
well-known. Originally discovered in conformal field theory ---
resulting from the Ward-identities as differential equations for
correlation functions in WZW models \cite{KZ} --- further appearance in
Chern-Simons theory \cite{CS}--\cite{FG} shows these equations to
represent underlying links between different physical areas in the
fundamental mathematical framework of affine algebras and quantum
groups.  In the context of geometrical quantization of Chern-Simons
theory \cite{CS,AWD,Hit}, KZ equations arise as conditions of
invariance of the quantization procedure with respect to a holomorphic
change of polarization.

In \cite{R} it was realized the link of KZ equations with a classical
mathematical object --- the Schlesinger equations on the Riemann
sphere \cite{S}.  Playing in particular an important role in the
theory of integrable systems \cite{J}, Schlesinger equations describe
monodromy preserving deformations of matrix differential equations
with rational coefficients on the sphere.  The procedure of canonical
quantization of the Schlesinger equations in the framework of
``multi-time'' Hamiltonian formalism derived in \cite{Jimbo1} leads
(in the Schr\"{o}dinger picture) to the KZ equations. Being rewritten
in the Heisenberg picture \cite{H} the KZ equations turn out to
coincide with the Schlesinger system for higher dimensional matrices
supplied with additional commutation relations.  It is interesting to
note that the Schlesinger equations, being the classical counterpart
of the KZ equations, in general case do not admit solutions in terms
of quadratures, although the KZ system does \cite{SV}.  In this sense
the quantized system (KZ) turns out to be simpler then the classical
one (Schlesinger).

In \cite{KN} a similar ``two-time'' quantization procedure was applied
to the stationary axisymmetric Einstein equations and allowed to
express the explicit solutions of related Wheeler-deWitt equations in
terms of the integral representation for solutions of KZ equations
found in \cite{SV}.\\

The main result of the present paper is the existence of a similar
link between KZ and the Schlesinger system on the torus. The KZ
equations on the torus --- usually referred to as
Knizhnik-Zamolodchikov-Bernard (KZB) equations --- were first derived
in the framework of conformal field theory on higher genus Riemann
surfaces \cite{B} and further studied by several authors (see \cite{F}
and references therein). We show them to arise from quantization of
isomonodromic quantization on the torus.  As a by-product of our
construction we derive the explicit form of the Schlesinger equations
on the torus, which also seems to be new.\footnote{In hidden form
these equations were exploited, for example, in the investigation of
the XYZ Landau-Lifshitz equations \cite{ItsBikbaev}, where the
associated spectral problem lives on the curve of genus one.}

The main differences to the case of the Riemann sphere are the
following:
\begin{itemize}
\item
The basic variables, meromorphic Lie-algebra valued one-forms, are
non-singlevalued (``twisted'') on the torus.  The origin of the twist
becomes clear, if one considers this one-form as the holomorphic gauge
of a general (single-valued) flat connection on the torus: the gauge
transformation which leads to holomorphic gauge turns out to be
non-invariant with respect to translation along the basic periods in
general.

In WZW models on the torus, twisted quantities were
introduced by hand to achieve completeness of the Ward identities by
defining a proper action of affine zero modes on character valued
correlation functions, which led to the KZB equations \cite{B}.

\item
In addition to the deformation with respect to the positions of the
poles of the connection on the torus, the Schlesinger system on the
torus contains a deformation equation with respect to the moduli
(i.e.~$b$-period) of the torus. Thus the set of the deformation
parameters coincides with the full set of coordinates on the moduli
space of punctured Riemann surfaces.
\end{itemize}

The multi-time Hamiltonian formulation of the genus one
Schlesinger equations with respect to these deformation parameters
reveals the link with Chern-Simons theory on the torus --- the relevant
Poisson bracket turns out to be the Dirac bracket of Chern-Simons
Poisson bracket \cite{CS} with respect to the flatness and gauge
fixing constraints.\footnote{The related fact on the sphere was
noticed without proof in \cite{FR}.}
Canonical quantization of this Hamiltonian formulation is shown to
give the KZB equations in the form of \cite{F}.

In Section 2 we discuss the links between all these objects for the
case of the sphere. In Section 3 we show how to extend this treatment on
the torus with the main focus on the features which arise due to
topological non-triviality. Section 4 contains some further comments.

\section{Holomorphic Poisson bracket and isomonodro\-mic quantization
on the sphere}

We consider the space of holomorphic Lie-algebra valued one-forms on
the punctured Riemann sphere, that are meromorphic with simple poles
on the whole sphere. These forms may be viewed as connections on a
trivial bundle. To simplify notation and without any loss of
generality, in the explicit expressions we restrict to the case of
${\mathfrak g}={\mathfrak su}(2)$.

Introducing local coordinates on the sphere by marking a point
$\infty$, an element $A(z) dz$ of this space is uniquely determined by
its poles $z_j$ and the corresponding residues $A_j$ taking values in
${\mathfrak g}$:
\be
A(z) = \sum_{j=1}^N \frac{A_j}{z-z_j}
\ee

Holomorphic behavior at infinity is ensured by
\be
Q := \sum_j A_j = 0 \label{constraint}
\ee

There is a natural Poisson structure on the space of holomorphic
connections on the punctured complex plane, that may be formulated in
the equivalent expressions:
\ba
\{A^a_i,A^b_j\} &=& 2\d_{ij}\e^{abc}A^c_i \label{bracket}\\
\Leftrightarrow \hspace{1.9em} \{A^a(z),A^b(w)\} &=& -\e^{abc}
\frac{A^c(z)-A^c(w)}{z-w} \non
\Leftrightarrow \qquad \{A(z)\stackrel{\otimes}{,}A(w)\} &=&
[r(z-w),A(z)\otimes I + I\otimes A(w)] \nn
\ea
with a classical $r$-matrix $r(z)=\frac{\Pi}{z}$, $\Pi$ being the
$4\times 4$ permutation operator.

The condition (\ref{constraint}) that restricts the connections to
live on the sphere, transforms as a first-class constraint under this
bracket: $\{Q^a,Q^b\} = \e^{abc}Q^c $\\

\subsection{Holomorphic bracket from gauge fixed Chern-Simons theory}

Let us shortly describe the relation of the bracket (\ref{bracket}) to
the fundamental Atiyah-Bott symplectic structure, that was claimed by
Fock and Rosly \cite{FR}.

The space of smooth connections on a Riemann surface is endowed with
the natural symplectic form \cite{AB}:
\ben
\W = \rm{tr}\int \delta A \wedge \delta A,
\een
that leads to the Poisson bracket
\be
\{A^a_z(z),A^b_{\bar{z}}(w)\} = \d^{ab}\delta^{(2)}(z-w), \label{ABbracket}
\ee
where the connection $A$ is split into $A_z dz + A_{\bar{z}} d\bar{z}$
and the $\delta$-function is understood as a real two-dimensional
$\delta$-function: $\delta^{(2)}(x+iy)\equiv\delta(x)\delta(y)$.

The condition of flatness is $F = dA + A\wedge A = 0$ and transforms as
a first-class constraint under the bracket:
\ben
\{F^a(z),F^b(w)\} = \e^{abc}F^c(z)\delta^{(2)}(z-w)
\een

This constraint generates gauge transformations
\be
A \mapsto gAg^{-1} + dg g^{-1} \label{gt}
\ee
that leave the symplectic structure invariant.\\

These brackets and constraints arise naturally from the Chern-Simons
 action \cite{CS}. They may be extended to punctured Riemann surfaces
 if the singularities of the connection restrict to first order poles,
 leading to $\delta$-function-like singularities of the curvature
 \cite{CS, CS2}.\\

We may now fix the gauge freedom by choosing the gauge $A_{\bar{z}}=0$
that makes flatness turn into holomorphy. Whereas there is certainly
no problem to achieve this gauge on the complex plane, in the case of
the punctured sphere a few comments are in order:
\begin{itemize}
\item{Poles in $A_{\bar{z}}$ may be removed by gauge transformations
of the local form $(z-z_j)^{A_j}(\bar{z}-\bar{z_j})^{A_j}$. Even
though these gauge transformations are singular in $z_j$ they should
yield no severe problems, as they do not change the residue of the
pole, but shift it into a pole of the surviving $(1,0)$-form $A_z
dz$.}
\item{A more subtle problem is the non-existence of appropriate global
gauge-transforma\-tions in general. The space of smooth
Lie-algebra-valued $(0,1)$-forms on the sphere is naturally isomorphic
to the space of holomorphic bundles, the pure gauge forms
corresponding to the trivial bundle.

This means, in general gauging away $A_{\bar{z}}$ may be interpreted
as $A_z dz$ becoming a connection on a nontrivial bundle on the
sphere. An additional degree of freedom, measuring the non-triviality
of the transition functions would have to be introduced, the
$\Z$-valued Chern number in the case of the sphere. We will work out
an example of only locally defined gauge transformations in the next
section on the torus.

However, as on the sphere the space of pure gauge $(0,1)$-forms builds
a dense subset in the space of smooth $(0,1)$-forms \cite{AB,GK}, we
restrict the following considerations to this subspace. A closer
investigation of this restriction determines the subspace of physical
Chern-Simons states \cite{GK,FG}. In our framework this identification
should be achieved in terms of the monodromy algebra of the connection
$A$, see last section.}
\end{itemize}

The bracket between constraints and gauge-fixing condition is of the
form:
\be
\{F^a(z),A^b_{\bar{z}}(w)\} = -\delta^{ab}\partial_{\bar{z}}\delta^{(2)}
(z-w) + \e^{abc}A^c_{\bar{z}}(z)\delta^{(2)}(z-w) \label{brcon}
\ee

Following the Dirac procedure of gauge fixing \cite{D}, the bracket
(\ref{ABbracket}) has to be modified to make it consistent with the
now second-class constraints. The matrix of brackets (\ref{brcon}) can
be inverted using $\partial_{\bar{z}}\frac1{z}=\delta^{(2)}(z)$ and
yields exactly the bracket (\ref{bracket}).  Notice that in the
framework of geometric quantization the variables $A_{\bar{z}}$ and
$A_z$ are, according to (\ref{ABbracket}), considered as canonically
conjugated coordinate and momentum, respectively.  After the
holomorphic gauge fixing the surviving variable $A(z)\equiv A_z(z)$
looks, according to (\ref{bracket}), more like a combination of
angular momenta.
\\

Note that because of the appearance of $\partial_{\bar{z}}$ in
(\ref{brcon}), the holomorphic part of the constraints $F^a(z)$
survives as a first-class constraint. As holomorphic functions on the
sphere are constants, the remaining flatness conditions become
\ben
\int F^a(z) dz d\bar{z} = \int \partial_{\bar{z}} A^a(z) dz d\bar{z} =
\sum_j A_j^a = Q^a
\een

The holomorphic bracket (\ref{bracket}) is therefore induced by
restricting the fundamental Poisson structure on the space of smooth
connections to the space of flat connections and fixing holomorphic
gauge. The first-class constraint (\ref{constraint}) ensuring $A(z)$
to live on the sphere, arises naturally as surviving flatness
condition, generating the constant gauge transformations.

\subsection{Hamiltonian formulation of isomonodromic deformation}

We now describe isomonodromic deformation on the sphere in terms of
the introduced holomorphic Poisson structure. Consider the system of
linear differential equations:
\be
\partial\Psi(z) = A(z)\Psi(z) \label{zsph}
\ee

As $A(z)$ has simple poles, the Lie-group-valued function $\Psi(z)$
lives on a covering of the punctured sphere. Let $\Psi$ be normalized
to $\Psi(\infty)=I$, thereby marking one of the points $\infty$ on
this covering. In the neighborhood of the points $ z_i$, the function
$\Psi$ is given by:
\be
\Psi(z) = G_i\Psi_i(z)(z-z_i)^{T_i}C_i \label{local}
\ee
with $\Psi_i(z)=I+{\cal O}(z-z_i)$ being holomorphic and
invertible. The relation to the residues of the connection is given by
$A_i = G_iT_iG_i^{-1}$.

The local behavior (\ref{local}) also yields explicit expressions for
the monodromies around the singularities:
\ben
\Psi(z) \mapsto \Psi(z)M_i  \qquad \mbox{for $z$ encircling $z_i$}
\een
with
\ben
M_i = C_i^{-1}\exp(2\pi i T_i)C_i
\een

Note that the normalization $\Psi(\infty)=I$ couples the freedom of
 right multiplication in the linear system (\ref{zsph}) to the left
 action of constant gauge transformations (\ref{gt}) on $\Psi$ to $
 \Psi \mapsto g\Psi g^{-1}$ and therefore $M_i \mapsto gM_i g^{-1}$\\

The aim of isomonodromic deformation \cite{J} is the investigation of
a family of linear systems (\ref{zsph}) parameterized by the choice of
singular points $z_i$, that have the same monodromies. In other words,
one studies the change of the connection data $A_i$ with respect to a
change in the parameters of the Riemann surface that is required to
keep the monodromy data constant. Treating $A(z)$ and $\Psi(z)$ as
functions of $z$ and $z_i$, these isomonodromy conditions impose a formal
condition of $z_i$-independence of the monodromy data $T_i$ and
$C_i$. This ensures the function $\partial_{z_i}\Psi\Psi^{-1}(z)$ to
have a simple pole in $z_i$:
\be
\partial_{z_i}\Psi(z) = \frac{-A^i}{z-z_i}\Psi(z) \label{zisph}
\ee

Compatibility of these equations with the system (\ref{zsph}) yields the
classical Schlesinger equations \cite{S}
\be
\partial_{z_i}A_j = \frac{[A_i,A_j]}{z_i-z_j} \quad
\mbox{for $j\not=i$}\quad,\qquad \partial_{z_i}A_i =
-\sum_{j\not=i}\frac{[A_i,A_j]}{z_i-z_j} \label{schles}
\ee

A Hamiltonian description of this dependence is given by \cite{Jimbo1}
\be
H_i = \sum_{j\not=i}\frac{{\rm tr}(A_iA_j)}{z_i-z_j}
\ee
in the described holomorphic Poisson-bracket (\ref{bracket}).\\

As was first noticed by Reshetikhin \cite{R} and realized in a simpler
form in \cite{H}, quantization of this system leads directly to the
Knizhnik-Zamolodchikov equations, that are known as differential
equations for correlation functions in conformal field theory
\cite{KZ}.

Quantization of the system is performed straight-forward by replacing
the Poisson structure by commutators. Shifting the $z_i$-dependence of
the operators $A_i^a$ (\ref{schles}) into the states on which these
operators act, corresponds to a transition from the Heisenberg picture
to the Schr\"odinger picture in ordinary quantum mechanics. In the
Schr\"odinger representation the quantum states $|\w\rangle$ then are
sections of a holomorphic $V\equiv\bigotimes_j V_j$ vector bundle over
$X_0\equiv\C^{N}\setminus\{\mbox{diagonal hyperplanes}\}$. The
$z_i$-independent operator-valued coordinates of $A_i$ are realized as
\be
A_i^a = i\hbar~I\otimes\dots\otimes t^a_i \otimes\dots\otimes I
\label{Aai}\ee
where $t^a_i$ acts in the representation $V_i$.

In this Schr\"odinger picture the quantum states $|\w\rangle$ obey the
following multi-time $z_i$-dynamics:
\be
\partial_{z_i} |\w\rangle = H_i |\w\rangle =
i\hbar \sum_{j\not=i}\frac{\W_{ij}}{z_i-z_j} |\w\rangle \label{kz}
\ee
Here, $\W_{ij}={\rm tr}(t_i\otimes t_j)$ is the Casimir operator
of the algebra acting non-trivially only on $V_i$ and $V_j$.

System (\ref{kz}) may be equivalently rewritten in the Heisenberg picture
introducing the multi-time evolution operator $U(\{z_i\})$ by
$$\partial_{z_i} U= H_i U\;\;\;,\;\;\; U(\{z_i=0\}) = {\rm id} $$

Then in terms of the variables
\be
\hat{A}_i^a\equiv U A_i^a U^{-1}
\label{st}
\ee
the quantum equations of motion give rise to higher-dimensional
Schlesinger equations with the matrix entries $A_i^a$ being operators
in $V$. These equations turn out to be a very special case of the
general $({\rm dim}{\mathfrak g}\times{\rm dim} V)$-dimensional
classical Schlesinger system since all operators $\hat{A}_i^a$ may be
simultaneously transformed to (\ref{Aai}) by some similarity
transformation (\ref{st}).

The system (\ref{kz}) that defines horizontal sections on the bundle
of quantum states is the famous Knizhnik-Zamolodchikov system
\cite{KZ} that arises here in a different context.

\section{Quantization of Schlesinger equations in genus one}

We are going to repeat the analysis of the last section for the case
of the torus now, which will show the link between isomonodromic
deformation and the KZB-equations on the torus. The conceptual novelty
of twisted functions, that is introduced more or less by hand in WZW
conformal field theories on the torus in order to get a proper
description of the action of inserted affine zero modes in the
correlation functions \cite{B}, enters the game in a very natural way
in our treatment.

\subsection{Holomorphic gauge fixing}

We start again from a smooth $\mathfrak{su}(2)$-valued one-form $A$ on
the torus. In the explicit formulae we will use standard Chevalley
generators $t^3, t^\pm$. Denote the periods of the torus by $1$ and
$\t$.

Holomorphic gauge $A_{\bar{z}}=0$ can not be achieved in
general. However, taking into account our remarks from the previous
section, the essential fact is \cite{FG}, that a dense subspace of
smooth (0,1)-forms can be gauged into constants of the form
\be
A_{\bar{z}}=\frac{2\pi i \l}{\t-\bar{\t}} \s_3~,\qquad \l\in \C
\label{gauge}
\ee

The holomorphic gauge condition would require an additional gauge
transformation of the kind $g=\exp(2\pi i \l
\frac{z-\bar{z}}{\t-\bar{\t}} \s_3)$. This is obviously multi-valued
on the torus, having a multiplicative twist: $g \mapsto \exp(2\pi i \l
\s_3 )g$ for $z$ encircling the fundamental $(0,\t)$-cycle. The result
of a gauge transformation of this kind is a twist in the remaining
holomorphic (1,0)-form $A(z)$:
\be
A(z+1) = A(z) \qquad A(z+\t) = e^{2\p i\l {\rm ad}\s_3}A(z) \label{twist}
\ee

In components this reads:
\ben
A^3(z+\t) = A^3(z) \qquad A^{\pm}(z+\t) = e^{\pm4\p i\l}A^{\pm}(z)
\een

Even though in general gauge transformations must be defined globally
single-valued in order to conserve physics, in this case our
proceeding is justified by the fact, that the non-gauge-trivial part
of $A_{\bar{z}}$ survives as an arising twist of the holomorphic
connection.

This is how the holomorphic gauge causes the appearance of twisted
quantities in a very natural way.\\

\subsection{Some meromorphic functions on the torus}

Before we start to investigate isomonodromic quantization on the
torus, let us collect some simple facts about twisted meromorphic
functions on the torus. A basic ingredient to describe functions of
this kind, is Jacobi's theta-function:
\ben
\th(z):=\sum_{n\in \Z}e^{2\pi i(\frac12 n^2\t+nz)},
\een
which is holomorphic, twisted as: $\th(z+1)=\th(z),\quad \th(z+\t)=
e^{-i\pi(\t+2z)}\th(z)$ and has simple zeros for
$z\in \half(\t+1)+\Z+\t \Z$.

Standard combinations are the functions \cite{F}:
\be
\r(z):= \frac{\th'(z-\frac12(\t+1))}{\th(z-\frac12(\t+1))}+i\pi
\quad\mbox{and}\quad \s_\l(z):=\frac{\th(\l-z-\frac12(\t+1))\th'
(\frac12(\t+1))}{\th(z+\frac12(\t+1))\th(\l-\frac12(\t+1))}
\ee
which have simple poles with normalized residue in $z=0$ and additive
and multiplicative twist respectively:
\ben
\r(z+\t)=\r(z)-2\pi i \quad;\qquad \s_\l(z+\t)=e^{2\pi i\l}\s_\l(z)
\een\smallskip

We should still mention the properties
\ben
\r(-z)= -\r(z) \quad;\qquad  \s_\l(z)=-\s_{-\l}(-z)
\een
and the identity
\ben
\partial_\l\s_\l(x-y) = \s_\l(x-y)\Big(\r(z-x)-\r(z-y)\Big) -
\s_{-\l}(z-x)\s_\l(z-y)
\een

They can be proved checking residues and twist properties. All the
following calculations rely on the fact, that meromorphic functions on
the torus with simple poles are uniquely determined by their residues
if they are multiplicatively twisted, whereas functions with additive
or vanishing twist are determined only up to constants. Note further
that in generic situation there are no holomorphic twisted functions
on the torus.

\subsection{Isomonodromic deformation}

Equipped with these tools we can now start to describe the twisted
meromorphic connection $A(z)$. Because of its twist properties
(\ref{twist}), $A(z)$ is of the form:
\ba
A^{\pm}(z) &=& \sum_j A^{\pm}_j \s_{\pm2\l}(z-z_j) \\
A^3(z) &=& \sum_j A^3_j \r(z-z_j) - B^3 \nn
\ea

Define again $\Psi$ by the linear system
\be
\partial\Psi(z) = A(z)\Psi(z) \label{ztor}
\ee

The function $\Psi$ will get monodromies $M_i$ and $M_{(0,1)}$ from
the right hand side, if $z$ encircles $z_i$ or the $(0,1)$ cycle of
the torus. If $z$ runs along the $(0,\t)$ cycle, $\Psi$ will exhibit
an additional left monodromy due to the twist of $ A$:
\be
\Psi(z) \mapsto e^{2\pi i \l\s_3} \Psi(z) M_{(0,\t)} \label{mon}
\ee

Under isomonodromic deformation we will understand the invariance of
the right hand side monodromy data under the change of the parameters
of the punctured torus, which are the singular points $z_i$ and the
period $\t$. The connection data in this case are the residues $A_i$,
the additive constant $B^3$ and the twist $\l$.

Let us first investigate their $z_i$-dependence. In addition to the
residues of $\partial_{z_i}\Psi\Psi^{-1}$ we have to determine its
twist around $(0,\t)$ from isomonodromy conditions. Equation
(\ref{mon}) yields:
\ben
\Big(\partial_{z_i}\Psi\Psi^{-1}\Big)(z) \mapsto e^{2\pi i \l
{\rm ~ad}\s_3}\Big(\partial_{z_i}\Psi\Psi^{-1}\Big)(z) + 2\pi i
\partial_{z_i}\l\s_3
\een

This determines the form of the $z_i$-dependence of $\Psi$ to:
\ba
\Big(\partial_{z_i}\Psi\Psi^{-1}\Big)^{\pm}(z) &=&
-A^{\pm}_i\s_{\pm2\l}(z-z_i) \label{zitor}\\
\Big(\partial_{z_i}\Psi\Psi^{-1}\Big)^3(z) &=& -A^3_i\r(z-z_i)+B_i^3\nn
\ea

and further on yields the $z_i$-dependence of the twist parameter $\l$:
\be
\partial_i\l = A_i^3
\ee\smallskip

We can now proceed as in the previous section. Compatibility of the
equations (\ref{ztor}) and (\ref{zitor}) implies the following
Schlesinger equations
on the torus:
\ba
\partial_{z_i}A_j^3 &=& -A_i^+A_j^-\s_{2\l}(z_j-z_i)+A_i^-A_j^+
\s_{-2\l}(z_j-z_i)\quad;\qquad j\not=i \label{schlestor}\\
\partial_{z_i}A_i^3  &=& \sum_{j\not=i}A_i^+A_j^-\s_{2\l}(z_j-z_i) -
\sum_{j\not=i}A_i^-A_j^+\s_{-2\l}(z_j-z_i) \non
&& \non
\partial_{z_i}A_j^{\pm} &=& \pm 2A_i^{\pm}A_j^3\s_{\pm2\l}(z_j-z_i)
\mp 2A^3_iA^{\pm}_j\r(z_j-z_i) \pm 2B_i^3A_j^{\pm} \quad;
\qquad j\not=i \non
\partial_{z_i}A_i^{\pm} &=& \pm 2\sum_{j\not=i}A_i^{\pm}A_j^3\r(z_i-z_j)
\mp 2\sum_{j\not=i}A^3_iA^{\pm}_j\s_{\pm2\l}(z_i-z_j) \mp 2B^3A_i^{\pm}
\pm 2B_i^3A_i^{\pm} \non
&&\non
\partial_i B^3 &=& \frac12\sum_{j\not=i}\Big(A_i^-A_j^+
\partial_\l\s_{2\l}(z_i-z_j) - A_i^+A_j^-\partial_\l\s_{2\l}(z_j-z_i)
\Big) \nn
\ea

and a curvature condition on the constants $B_i$:
\be
\partial_i B^3_j-\partial_jB^3_i = \half A_i^-A_j^+
\partial_\l\s_{2\l}(z_i-z_j) - \half A_i^+A_j^-
\partial_\l\s_{2\l}(z_j-z_i) \label{curvature}
\ee\\

As on the sphere, it is possible to formulate this dependence as a
multi-time Hamiltonian structure. The Hamiltonians can be written down as
\ba
H_i &=& \sum_{j\not=i} \Big(2 A_i^3A_j^3\r(z_i-z_j) +
A_i^+A_j^-\s_{-2\l}(z_i-z_j) + A_i^-A_j^+\s_{2\l}(z_i-z_j) \Big) \non
&& {}- 2B^3A_i^3 + 2B_i^3\sum_j A_j^3 \label{hamtor}
\ea
in the Poisson structure:
\ba
\{A_i^a,A_j^b\} &=& 2\d_{ij} \e^{abc} A_i^c \label{brackettor}\\
\{\l,B^3\} &=& \half \nn
\ea

We strongly suppose that this structure arises from holomorphic
gauge-fixing of the original bracket (\ref{ABbracket}) in the same
way, as does the bracket (\ref{bracket}) on the sphere. In particular,
remembering the origin of $\l$ (\ref{gauge}), the second equation may
be viewed as a reminiscent of (\ref{ABbracket}) for the constant terms
of $A_z$ and $A_{\bar{z}}$.

In analogy with (\ref{bracket}) this Poisson structure admits a
generalized $r$-matrix formulation
\be
\{A(z)\stackrel{\otimes}{,}A(w)\} = [r(z-w),A(z)\otimes I +
I\otimes A(w)] - \partial_\l r(z-w) \Big(\sum_j A_j^3\Big)
\ee
with the twisted $r$-matrix
\be
r(z) = \half \r(z)(t^3\otimes t^3)+\s_{2\l}(z)(t^+\otimes t^-) +
\s_{-2\l}(z)(t^-\otimes t^+)
\label{r-matrix}
\ee
that in some sense restricts to a classical $r$-matrix formulation on
the constraint surface (\ref{constrainttor}) below. Validity of the
Jacobi identities is expressed by a twisted version of the classical
Yang-Baxter equation.  Notice that the twisted $r$-matrix
(\ref{r-matrix}) reminds the elliptic $r$-matrix arising in the
Hamiltonian formulation of the XYZ Landau-Lifshitz equation
\cite{Faddeev}, which, however, has different twist properties and
satisfies standard (non-twisted) Yang-Baxter equation.
\\

In addition to (\ref{schlestor}), compatibility of the linear systems
implies
\be
\sum_j A_j^3 = 0 \label{constrainttor}
\ee

In the holomorphic Poisson-structure this simply generates the
remaining constant gauge transformations compatible with
(\ref{gauge}). In contrast to the sphere, the gauge has been fixed
more rigorously on the torus in order to diagonalize the twist around
the $(0,\t)$-cycle. Ref.\,\cite{I} treats the weaker gauge of
arbitrary twist that accordingly leads to a stronger constraint. In
this gauge it is still possible to achieve a normalization
$\Psi(z_0)=I$ of $z_i$-independence of $\Psi$ at a fixed point $z_0$
like on the sphere, which turns out to be inconsistent with our
restrictive choice of gauge.

A consequence of this restrictive gauge on the torus is the appearance
of the undetermined constants $B_i$ in the Schlesinger equations
(\ref{schlestor}) on which we still have to spend some comments. The
way they arise in the Hamiltonians (\ref{hamtor}) proves them to
generate gauge transformations. This suggests to simply skip these
terms from the Hamiltonians, as is in fact done in the sequel, leading
to the KZB equations. However, this is certainly not compatible with
the curvature condition (\ref{curvature}) which in turn implies that
these truncated Hamiltonians only commute up to (\ref{constrainttor}),
meaning that the generated $z_i$ dynamics of the connection data
produces isomonodromic deformation only up to certain shifts in the
gauge orbit.  Nevertheless, this seems to be the most elegant way to
treat these gauge constants.\\

We will finally study isomonodromic deformation with respect to a
 change in the period $\t$ of the torus. This can be done in complete
 analogy with the just treated case. From (\ref{mon}) the twist of
 $\partial_\t\Psi\Psi^{-1}$ around $(0,\t)$ turns out to be
\ben
\Big(\partial_\t\Psi\Psi^{-1}\Big)(z) \mapsto e^{2\pi i \l
{\rm ~ad}\s_3}\Big(\partial_\t\Psi\Psi^{-1}\Big)(z) - e^{2\pi i \l
{\rm ~ad}\s_3}A(z) + 2\pi i \partial_\t \l \s_3
\een

which leads to the following $\t$-dependence of the function $\Psi$:
\ba
2\pi i\Big(\partial_\t\Psi\Psi^{-1}\Big)^{\pm}(z) &=& \mp
\frac12\sum_j A_j^{\pm}\partial_\l\s_{\pm2\l}(z-z_j) \label{tautor}\\
2\pi i\Big(\partial_\t\Psi\Psi^{-1}\Big)^3(z) &=& \frac12\sum_j A_j^3
\Big(\r(z-z_j)^2-\wp(z-z_j)\Big) +  B_\t^3 \nn
\ea
and determines the $\t$-dependence of the twist parameter
\be
\partial_\t \l = -\tpi B^3
\ee\smallskip

Compatibility of (\ref{ztor}) and (\ref{tautor}) now yields additional
Schlesinger-type equations:
\ba
2\pi i \partial_\t A_i^3 &=& -\frac12\sum_j A_i^+A_j^-
\partial_\l\s_{-2\l}(z_i-z_j)-\frac12\sum_j A_i^-A_j^+
\partial_\l\s_{2\l}(z_i-z_j) \label{schlestort}\\
2\pi i \partial_\t A_i^{\pm} &=&  \pm  \sum_j A_i^{\pm}A_j^3
\Big(\r(z_i-z_j)^2-\wp(z_i-z_j)\Big) \non
&& {} + \sum_j A_i^3A_j^{\pm} \partial_\l\s_{\pm2\l}(z_i-z_j) \pm
2B^3_\t A_i^ {\pm} \non
&&\non
2\pi i \partial_\t B^3 &=& -\frac18\sum_{i,j}\Big(A_i^+A_j^-
\partial^2_\l\s_{-2\l}(z_i-z_j)-A_i^-A_j^+
\partial^2_\l\s_{2\l}(z_i-z_j)\Big) \nn
\ea
together with a curvature condition for $\tpi\partial_{z_i} B^3_\t -
\partial_\t B_i^3$.\\

This dependence is described by the Hamiltonian
\ba
2\pi i\, H_\t &=&  \frac12\sum_{i,j}A_i^3A_j^3\Big(\r(z_i-z_j)^2-
\wp(z_i-z_j)\Big) \\
&&{} +\frac14 \sum_{i\not= j}\Big(A_i^+A_j^-
\partial_\l\s_{-2\l}(z_i-z_j)-A_i^-A_j^+
\partial_\l\s_{2\l}(z_i-z_j)\Big) + B^3B^3 + 2B_\t^3 \sum_j A_j^3 \nn
\ea
where again we will skip $B^3_\t$ under the above remarks.

\subsection{KZB-equations}

Quantization is again performed straightforward with (\ref{brackettor})
being replaced by
\ba
[A_i^a,A_j^b] &=& 2 i\hbar \d_{ij} \e^{abc} A_i^c \non
\,[\l,B^3] &=& \half i\hbar \nn
\ea

In the $z_i$-independent Schr\"odinger representation of the operators
they can be realized as:
\ba
A_i^a &=& i\hbar~I\otimes\dots\otimes t^a_i \otimes\dots\otimes I \non
B^3 &=& - \half i\hbar \partial_\l \nn
\ea
acting on quantum states $|\w\rangle$ that are $\l$-dependent sections
of a $V\equiv\bigotimes_j V_j$ bundle over\\
$X_1\equiv\{\mbox{fundamental domain of $\t$}\} \otimes
\C^{N}\setminus\{\mbox{diagonal hyperplanes}\}$.

The quantization of (\ref{schlestor}) and (\ref{schlestort}) in the
Schr\"odinger picture provides this bundle with the horizontal
connection:
\ba
\partial_{z_i} |\w\rangle ~=~ H_i \,|\w\rangle &=& \frac12i\hbar
t^3_i\partial_\l\,|\w\rangle + i\hbar\sum_{j\not=i}
\W^z_{ij}(z_i-z_j,\t,\l)\,|\w\rangle \label{kzb}\\
2\pi i \partial_{\t} |\w\rangle ~=~ 2\pi i H_\t \,|\w\rangle &=&
\frac14 i\hbar \partial^2_\l\,|\w\rangle + i\hbar\sum_{i,j}
\W^\t_{ij}(z_i-z_j,\t,\l)\,|\w\rangle \nn
\ea

with
\ba
\W^z_{ij}(z,\t,\l) &=& \frac12\r(z) (t_i^3\otimes t_j^3) +
\s_{-2\l}(z) (t_i^+\otimes t_j^-) + \s_{2\l}(z)(t_i^-\otimes t_j^+)
\non
\W^\t_{ij}(z,\t,\l) &=& \frac18\Big(\r(z)^2-\wp(z)\Big)
(t_i^3\otimes t_j^3) + \frac14\partial_\l\s_{-2\l}(z)
(t_i^+\otimes t_j^-) - \frac14\partial_\l\s_{2\l}(z)
(t_i^-\otimes t_j^+) \nn
\ea
acting non-trivially only on $V_i$ and $V_j$.\smallskip

This is the KZB connection, found in Ref.\,\cite{B} as differential
equations for character-valued correlation functions. The form
(\ref{kzb}) coincides exactly with the form presented in
Ref.\,\cite{F}. In particular, the term that includes the derivative
with respect to the twist parameter $\l$ is the explicit analogue of
the action of affine zero modes on correlation functions in WZW
models. We have thereby proved the fundamental role of the KZB
connection in the context of isomonodromic deformation on the torus.

We stress again that in contrast to the KZ-system on the sphere these
Hamiltonians only commute up to the constraint (\ref{constrainttor})
which implies the fact that the KZB-connection is flat only as a
connection on the subbundle of states annihilated by $\sum_j t_j^3$,
see \cite{F}.

\section{Concluding Remarks}
We have studied isomonodromic deformation on the torus and shown that
quantization of the Schlesin\-ger system on the torus leads to the KZB
system of differential equations. The result suggests further that the
quantization procedure of isomonodromic deformations on higher genus
Riemann surfaces should lead to the corresponding higher KZB equations
\cite{B,I}.

Let us close with some general remarks.

\begin{itemize}

\item
There obviously seem to be two fundamental links between quantization
of Chern-Simons theory and the system of KZ(B) equations, that arise
from geometric quantization \cite{AWD,Hit,O} and isomonodromic
quantization, respectively. The idea may be roughly sketched as
follows:

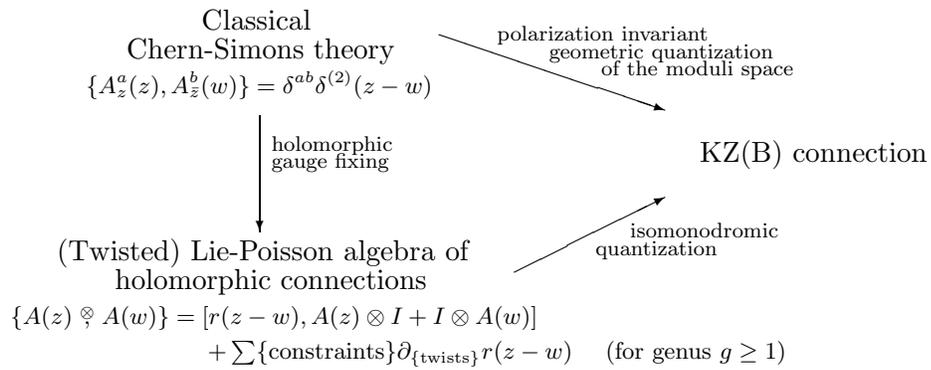
\begin{center}
 \setlength{\unitlength}{0.2em}
 \begin{picture}(150,74)
 \put(33,60){Classical}
 \put(20,55){Chern-Simons theory}
 \put(13,49){\footnotesize $\{A^a_z(z),A^b_{\bar{z}}(w)\} =
\d^{ab}\delta^{(2)}(z-w)$}
 \put(43,45){\vector(0,-1){20}}
 \put(45,39){\scriptsize holomorphic}
 \put(45,36){\scriptsize gauge fixing}
 \put(8,20){(Twisted) Lie-Poisson algebra of}
 \put(18,15){holomorphic connections}
 \put(0,9){\footnotesize $\{A(z)\stackrel{\otimes}{,}A(w)\} =
[r(z-w),A(z)\otimes I + I\otimes A(w)]$}
 \put(34,3){\footnotesize $+ \sum \{\rm constraints\}
\partial_{\rm \{twists\}} $$r(z-w)\quad$ (for genus $g\ge1$)}
 \put(119,37){KZ(B) connection}
 \put(74,59){\vector(3,-1){39}}
 \put(87,18){\vector(2,1){26}}
 \put(84,58){\scriptsize polarization invariant}
 \put(93,55){\scriptsize geometric quantization}
 \put(102,52){\scriptsize of the moduli space}
 \put(107,24){\scriptsize isomonodromic}
 \put(101,21){\scriptsize quantization}
 \end{picture}
\end{center}

\vspace{0.6cm}

\item
As both approaches lead directly to the KZ(B) connection, preserving
some structure on the bundle of quantum states, they should be
regarded as fundamental and probably somehow equivalent. In
particular, one is tempted to consider further properties of KZ(B)
equations, that were discovered in \cite{GK} and \cite{FG} --- namely
that these connections preserve the space of physical Chern-Simons
states by preserving a certain fusion-rule-like condition --- to be a
corollary of this more fundamental structure, the more, as the KZ(B)
systems prove to be sufficient but by no means unique in that
context.
\item
The physical subspace of Chern-Simons theory may be isolated in the
framework of isomonodromic quantization in the same way as in the
combinatorial quantization of Chern-Simons theory \cite{SA}, which we
have not depicted in the diagram. The basic variables here are the
monodromies of the connection $A$; their quantized algebra then turns
out to be a certain quantum group. As this approach yields the same
dimensions (Verlinde indices) of the physical spaces, on which these
operators are nontrivially represented, that are known from
geometrical quantization, it seems to be essentially equivalent. The
link to KZ(B) equations in this framework should be provided by their
appearance in the theory of quantum groups \cite{DV}.

\item
In spite of the close relation between the Poisson structures,
classically the Chern-Simons and Schlesinger systems are essentially
different: the dynamics is trivial in the first case and non-trivial
in the second one.  ``Times'' of the Schlesinger system on a Riemann
surface may be chosen as coordinates of the the moduli space of
punctured Riemann surfaces which play in (quantum) Chern-Simons theory
the role of formal deformation parameters. The natural question is:
what is the role of Schlesinger system in {\it classical} Chern-Simons
theory?

In this context one also has to notice an obvious link to 2+1 gravity:
in the approach of Witten \cite{Witten} 2+1 gravity is treated as a
Chern-Simons theory, which suggests relevance of Schlesinger equations
in this context; this agrees with the results of \cite{Welling} where
Schlesinger equations arise directly in $2+1$ gravity treated in the
framework of canonical Arnowitt-Deser-Misner formulation.

\end{itemize}

\paragraph{Acknowledgments}
We thank V.\,Schomerus and J.\,Teschner for useful discussions.
D.\,K.~is supported by Deutsche Forschungsgemeinschaft. H.\,S.~thanks
Studienstiftung des deutschen Volkes for financial support.

\end{document}